\documentclass[a4paper,twocolumn,english,aps,prl,superscriptaddress]{revtex4-1}
\usepackage[T1]{fontenc}
\usepackage[latin9]{inputenc}
\usepackage{color}
\usepackage{babel}
\usepackage{amsmath}
\usepackage{amssymb}
\usepackage{graphicx}
\DeclareGraphicsExtensions{.pdf}
\usepackage{amsmath,amssymb,bbold,bm,color}
\usepackage{float}
\usepackage{epstopdf}
\usepackage{tabularx}
\usepackage{braket}
\usepackage{booktabs}
\usepackage{array}
\usepackage{blindtext}
\usepackage{natbib}
\usepackage[unicode=true,pdfusetitle,
 bookmarks=true,bookmarksnumbered=false,bookmarksopen=false,
 breaklinks=false,pdfborder={0 0 1},backref=false,colorlinks=false]
 {hyperref}

\makeatletter

\usepackage{babel}
\usepackage{babel}

\begin{document}


\title{Dynamical signatures of Hardcore-Boson Supersolid on the Triangular Lattice}

\author{Xingchuan Zhu}
\thanks{These authors contributed equally to this work.}
\affiliation{School of Intelligence Science and Technology, Nanjing University of Science and Technology, Nanjing 210094, China}

\author{Qingyang Guo}
\thanks{These authors contributed equally to this work.}
\affiliation{School of Physics, Beihang University,
	Beijing, 100191, China}

\author{Junsong Sun}
\thanks{These authors contributed equally to this work.}
\affiliation{School of Physics, Beihang University,
	Beijing, 100191, China}

\author{Yujun Fu}
\affiliation{School of Physics, Beihang University,
	Beijing, 100191, China}

\author{Xingmin Huo}
\affiliation{School of Physics, Beihang University,
	Beijing, 100191, China}

\author{Yixuan Huang}
\affiliation{RIKEN Center for Emergent Matter Science (CEMS), Wako 351-0198, Japan}

\author{Shiping Feng}
\affiliation{Department of Physics, Beijing Normal University, Zhuhai 519087, China and School of Physics and Astronomy, Beijing Normal University, Beijing 100875, China}

\author{Richard T. Scalettar}
\affiliation{Physics Department, University of California, Davis, CA 95616, USA}

\author{Huaiming Guo}
\email{hmguo@buaa.edu.cn}
\affiliation{School of Physics, Beihang University,
Beijing, 100191, China}

\begin{abstract}
We investigate the dynamical signatures of the supersolid phase in the triangular-lattice hardcore Bose-Hubbard model via large-scale quantum Monte Carlo simulations and linear spin-wave theory. We reveal a distinct momentum-space separation in the spectral weights. The transverse spectrum shows gapless Goldstone modes at the $\Gamma$ and $K$ points, while the longitudinal spectrum features a gapless mode at $K$ and a roton-like minimum at $M$. Directly in the imaginary-time domain, the supersolid is unambiguously characterized by the coexistence of transverse spectral lines at $\Gamma$ and $K$. Furthermore, we show that strong quantum fluctuations induce a breakdown of the sharp quasiparticle picture, leading to asymmetric line shapes and broad continua. 
Our work establishes the definitive dynamical signatures of the hardcore-boson supersolid, providing key insights into  analogous spin supersolids in triangular quantum magnets.
\end{abstract}


\date{\today}

\maketitle


\textit{\textcolor{blue}{Introduction.--}}
The supersolid (SS) state, a quantum phase that simultaneously breaks spatial translation and global $U(1)$ gauge symmetries, has evolved from a long-sought but elusive goal in $^4$He~\cite{Andreev1969,Leggett1970,Chester1970,Kim2004,Boninsegni2012,Kim2012} to an experimentally realized state in dipolar quantum gases~\cite{Tanzi2019,Chomaz2019,Li2017,Leonard2017,Tanzi2019b,Norcia2021,casotti2024observation,Bland2022,Cinti2010,guo2019low,Birth2021,tanzi2021evidence} and, more recently, in quantum magnets~\cite{Gao2022,Xiang2024Giant,huang2026emergentspinsupersolidsfrustrated,shu2026giant}. In the latter context, the Matsubara--Matsuda transformation maps spin-$1/2$ degrees of freedom onto hardcore bosons~\cite{Matsubara1956}, establishing a direct correspondence where the SS phase is identified in triangular-lattice antiferromagnets with easy-axis anisotropy~\cite{Jiang2009,PhysRevLett.100.090401,Wang2009,Heidarian2010,Yamamoto2014,Sellmann2015,Starykh2015,PhysRevLett.134.196702,PhysRevB.62.15067,PhysRevLett.99.217205,PhysRevLett.131.116702,PhysRevLett.98.227201}. Recent experiments on candidate materials such as Na$_2$BaCo(PO$_4$)$_2$ and K$_2\text{Co}(\text{SeO}_3)_2$ have provided compelling thermodynamic and diffraction evidence for such spin SS states~\cite{Zhong2019,Li2020,Lee2021,gsk8-1k9q,cui2025spinsupersolidityinducedquantumcriticality,Chen2026Supersolid}.

However, the dynamical signatures of this phase remain elusive. Inelastic neutron-scattering measurements reveal spectra dominated by a broad continuum and intricate mode structures-including gapless Goldstone branches and roton-like minima-that challenge conventional linear spin-wave-theory (LSWT) descriptions~\cite{Zhu2024Wannier,Sheng2025Continuum,Zhu2024Wannier,Zhu2025Wannier,Gao2024Double}. Whether these features are intrinsic to generic SS phases or instead arise from material-specific effects remains an open question. Addressing this issue is theoretically demanding~\cite{Chi2024Dynamical,xdwz-8k94,PhysRevB.111.104435,PhysRevResearch.6.033031,PhysRevB.111.L060402,mauri2025nonlinearspinwavetheory,PhysRevB.111.174442,PhysRevB.98.184403,PhysRevB.88.094407,jtjk-x2lw}, as simulating frustrated spin models hosting a spin SS is hindered by the minus-sign problem in quantum Monte Carlo (QMC) simulations and the restriction to quasi-1D geometries in density-matrix renormalization group calculations.


In this manuscript, we investigate the dynamical signatures of the SS phase based on the hardcore Bose-Hubbard model on a triangular lattice. Although the equivalent spin model belongs to the generic XXZ Heisenberg family, it occupies a distinct parameter regime from the XXZ antiferromagnetic (AF) models that host experimentally realized spin SS phases. Notably, while the bosonic SS phase remains continuously connected to its spin counterpart in the full parameter space, the absence of the sign problem in the Bose-Hubbard model permits unbiased, large-scale QMC simulations. We calculate the dynamical correlations of the system by combining LSWT and QMC simulations. Specifically, QMC directly yields imaginary-time correlation functions, and the dynamical structure factor (DSF) is obtained via numerical analytic continuation. By analyzing the QMC data in both the imaginary-time and real-frequency domains, complemented by LSWT, we consistently identify the intrinsic dynamical spectral properties of the SS phase.

\begin{figure}[htp]
    \centering
    \includegraphics[width=0.49\textwidth]{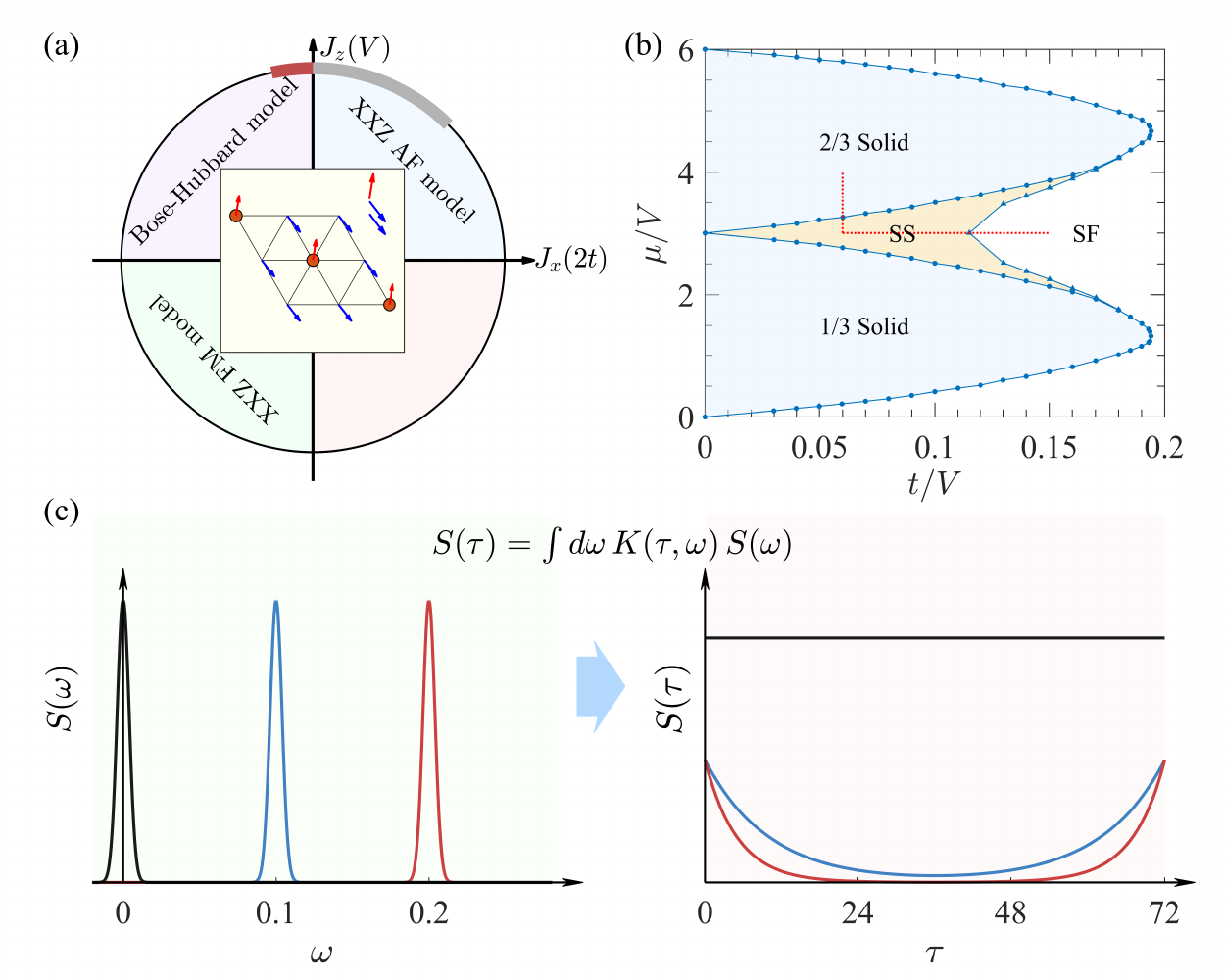} 
    \caption{(a) Distribution of supersolid phases across the full parameter space of the XXZ Heisenberg model. The inset shows the classical spin configuration of the hardcore-boson supersolid.  (b) Phase diagram of the hardcore Bose-Hubbard model in the ($t/V$, $\mu/V$) plane, including the solids at fillings $1/3$ and $2/3$, as well as supersolid and superfluid phases. The red cuts mark the phase-transition path along which we investigate the evolution of the dynamical correlations. (c) Qualitative correspondence between imaginary-time and real-frequency dynamical correlations, using $S(\omega)=\delta(\omega-\omega_0)$. Here, $0\le\tau\le72$ is a typical range for our simulatons at $L=36$ and $\beta=2 L$ }
    \label{fig1}
\end{figure}

\textit{\textcolor{blue}{Model and Method.--}}
We study the hardcore Bose-Hubbard model on a triangular lattice, equivalently described by the spin-$1/2$ XXZ Hamiltonian:
\begin{equation}
    H = \sum_{\langle i,j \rangle} \left[ -2t (S_i^x S_j^x + S_i^y S_j^y) + V S_i^z S_j^z \right] - h \sum_i S_i^z.
    \label{eq:hamiltonian}
\end{equation}
Here, the boson hopping $t$ corresponds to ferromagnetic (FM) XY exchange, the nearest-neighbor repulsion $V$ maps to AF Ising coupling, and the effective field $h=\mu - 3V$ is controlled by the chemical potential $\mu$, with $h=0$ corresponding to $\langle S_z\rangle=0$, i.e., the half-filling of hardcore bosons. In the generic XXZ Hamiltonian, the Bose-Hubbard model represents a specific parameter regime in the second quadrant [see Fig.\ref{fig1}~(a)]. The interplay of geometric frustration and interaction anisotropy ($t/V$) yields a rich phase diagram featuring superfluid, $\sqrt{3}\times\sqrt{3}$ solid, and supersolid phases~\cite{Wessel2005,Melko2005,Heidarian2005,Boninsegni2005,Sengupta2005,Batrouni2006,Batrouni2000,PhysRevB.75.094501,PhysRevB.72.134502,PhysRevB.74.214517,PhysRevLett.100.147204,PhysRevB.84.174515,landig2016quantum}. The latter, emerging upon doping the solid, is characterized by coexisting diagonal density-wave order and off-diagonal superfluidity. 


We study the model in Eq.~(\ref{eq:hamiltonian}) using LSWT followed by stochastic series expansion (SSE) QMC with directed loop
updates\cite{PhysRevB.43.5950,PhysRevE.66.046701,PhysRevE.67.046701,sandvik2010computational,PhysRevE.71.036706,PhysRevE.64.066701}. The SSE method expands the partition function in power series, mapping the quantum system to a classical configuration of operator strings and spin states.
Unlike path-integral formulations, this algorithm is free from systematic Trotter errors. The directed loop updates allow for global changes in the operator strings and spin configurations, enabling highly efficient sampling of the configuration space. Crucially, despite the triangular geometry, the FM nature of the in-plane coupling eliminates the sign problem, making unbiased numerical simulations feasible.
We compute the DSF~$S^{\alpha\beta}(\mathbf{k}, \omega)$ by analytically continuing the imaginary-time QMC data~$S^{\alpha\beta}(\mathbf{k}, \tau) = \langle S^\alpha_\mathbf{k}(\tau) S^\beta_{-\mathbf{k}}(0) \rangle$. We employ the stochastic maximum entropy method to resolve both longitudinal ($S_{\parallel} = S^{zz}$) and transverse ($S_{\perp} = S^{xx} + S^{yy}$) components~\cite{10.21468/SciPostPhysCodeb.1-v2.4,beach2004identifyingmaximumentropymethod}. Our QMC simulations are performed on linear lattice sizes up to~$L=36$, with the inverse temperature scaled as~$\beta=2L$ to ensure convergence to the ground state.

\textit{\textcolor{blue}{Correspondence between imaginary-time and real-frequency dynamical signatures.--}}
Given the relation ~$S^{\alpha\beta}(\mathbf{k}, \tau) = \int_{0}^{\infty} d\omega K(\tau, \omega) S^{\alpha\beta}(\mathbf{k}, \omega)$ with the finite-temperature bosonic kernel
~$K(\tau, \omega) = e^{-\tau\omega} + e^{-(\beta-\tau)\omega}$, extracting $S^{\alpha\beta}(\mathbf{k},\omega)$ from $S^{\alpha\beta}(\mathbf{k},\tau)$ remains an inherently ill-posed inverse problem. Because both representations encode the same dynamical information, it is instructive to analyze the signatures directly in imaginary time, where specific features may appear more distinctly. 

A qualitative understanding of the correspondence between imaginary-time and real-frequency dynamical correlations can be obtained by considering a single-mode spectrum $S(\mathbf{k},\omega)=\delta(\omega-\omega_0)$, for which $S(\mathbf{k},\tau)=K(\tau,\omega_0)$. The resulting $S(\mathbf{k},\tau)$ is symmetric about $\beta/2$ and exhibits exponential decay with $\tau$ in each half of the interval $[0,\beta]$, except for the special case $\omega_0=0$, where it remains constant~[see Fig.~\ref{fig1}(c)]. The decay rate is determined by the nonzero energy $\omega_0$, becoming faster for larger values of $\omega_0$. 

Basically, the QMC-calculated $S(\mathbf{k},\tau)$ exhibits two archetypal behaviors, which can be qualitatively accounted for by the above analysis.
A nearly constant, high-magnitude profile, typically observed at $K$ and $\Gamma$, signals the presence of long-range order and corresponds to a dominant spectral weight at $\omega=0$. In contrast, a rapidly decaying $S(\mathbf{k},\tau)$ with suppressed intensity, which either vanishes or reaches a small residual value at large $\tau$, reflects finite-energy dispersive excitations.

\begin{figure}[htp]
    \centering
    \includegraphics[width=0.5\textwidth]{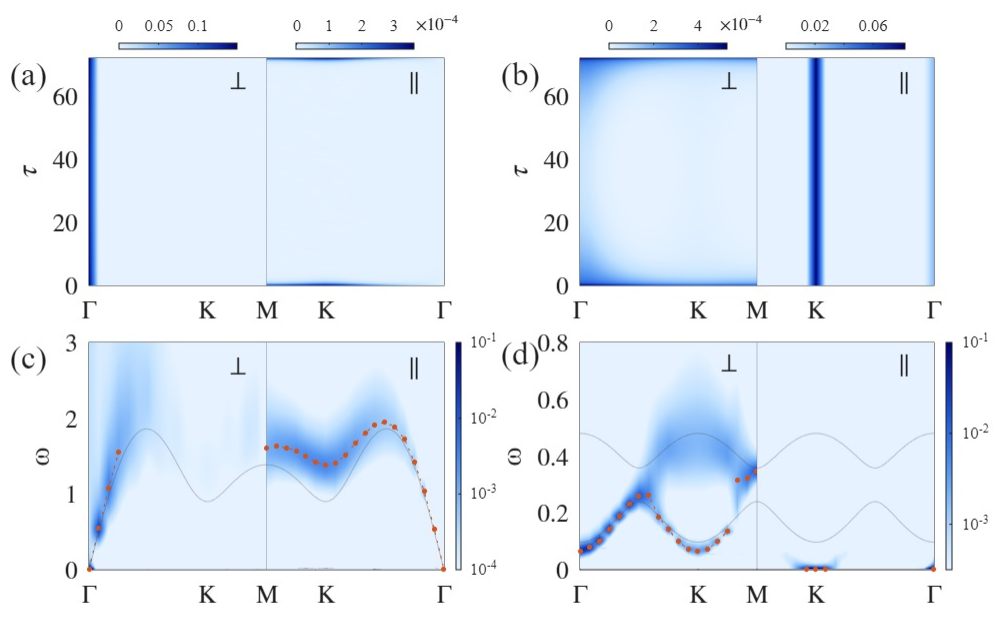} 
    \caption{Imaginary-time DSF obtained from QMC simulations for (a) the superfluid phase with $\rho=0.5$ at $t/V=0.3,\, \mu/V=3$ and (b) the solid phase with $\rho=2/3$ at $t/V=0.06,\, \mu/V=3.3$. (c) and (d) show the corresponding real-frequency DSFs obtained via numerical analytic continuation. The red solid circles in (c) and (d) represent the peak positions of the QMC spectra, while the black lines denote the magnon dispersions calculated from LSWT.}
    \label{fig2}
\end{figure}

\textit{\textcolor{blue}{Dynamical correlations in the Superfluid and Solid Phases.--}} 
To establish a benchmark for identifying the unique dynamical signatures of the supersolid state, we first illustrate the imaginary-time correlations and DSFs of the pure superfluid and solid phases in Fig.~\ref{fig2}. For the superfluid at $t/V=0.3$ and $\mu/V=3$, the imaginary-time correlations of the different components exhibit distinct distributions across momentum space. Accordingly, the spectral weight of $S^{\perp}(\mathbf{k},\omega)$ concentrated near the $\Gamma$ point, combined with $S^{\parallel}(\mathbf{k},\omega)$ distributed along the $K$--$M$ path, constitutes a single excitation branch. The distinct momentum distributions of the spectral weights in $S^{\parallel}(\mathbf{k},\omega)$ and $S^{\perp}(\mathbf{k},\omega)$ 
can be qualitatively understood within the framework of LSWT [See Supplemental Material (SM)\cite{SM}]. Furthermore, while LSWT accurately reproduces the sharp $S^{\perp}$ dispersion near the $\Gamma$ point, deviations arise along the $K$--$M$ path, where the LSWT prediction merely traces the lower-energy boundary of the broadened spectrum.


A defining signature of the superfluid phase is the gapless Goldstone mode emerging at the 
$\Gamma$ point, driven by the FM exchange coupling in the 
$xy$ plane. In the imaginary-time domain, this uniform macroscopic order manifests as a pronounced spectral line in $S^{\perp}(\mathbf{k},\tau)$ precisely at 
$\Gamma$. In its immediate vicinity, the spectral weight exhibits a rapid decay with 
$\tau$, directly characterizing the linear dispersion of the Goldstone excitations. 


In the solid phase at $t/V=0.06$ and $\mu/V = 3.3$, the up-up-down magnetic order manifests as a pronounced signal in $S_{\parallel}(\mathbf{k},\tau)$ at the $K$ point and a much weaker one at $\Gamma$, while remaining negligible at other momenta. The corresponding spectral weight of $S_{\parallel}(\mathbf{k},\omega)$ is concentrated exclusively at $\omega = 0$ at both the $K$ and $\Gamma$ points. The transverse component, $S_{\perp}(\mathbf{k},\tau)$, exhibits a rapid decay and is predominantly concentrated near $\tau = 0(\beta)$, with a decay profile that varies with momentum. The extracted $S_{\perp}(\mathbf{k},\omega)$ features a sharp, well-defined gapped low-energy dispersion that closely tracks the LSWT prediction, except near the $M$ point, where the spectral weight shifts to higher energies. As in the superfluid phase, the dominant spectral weight resides near the $\Gamma$ point, though it now acquires a finite gap.



\begin{figure}[htp]
    \centering
    \includegraphics[width=0.5\textwidth]{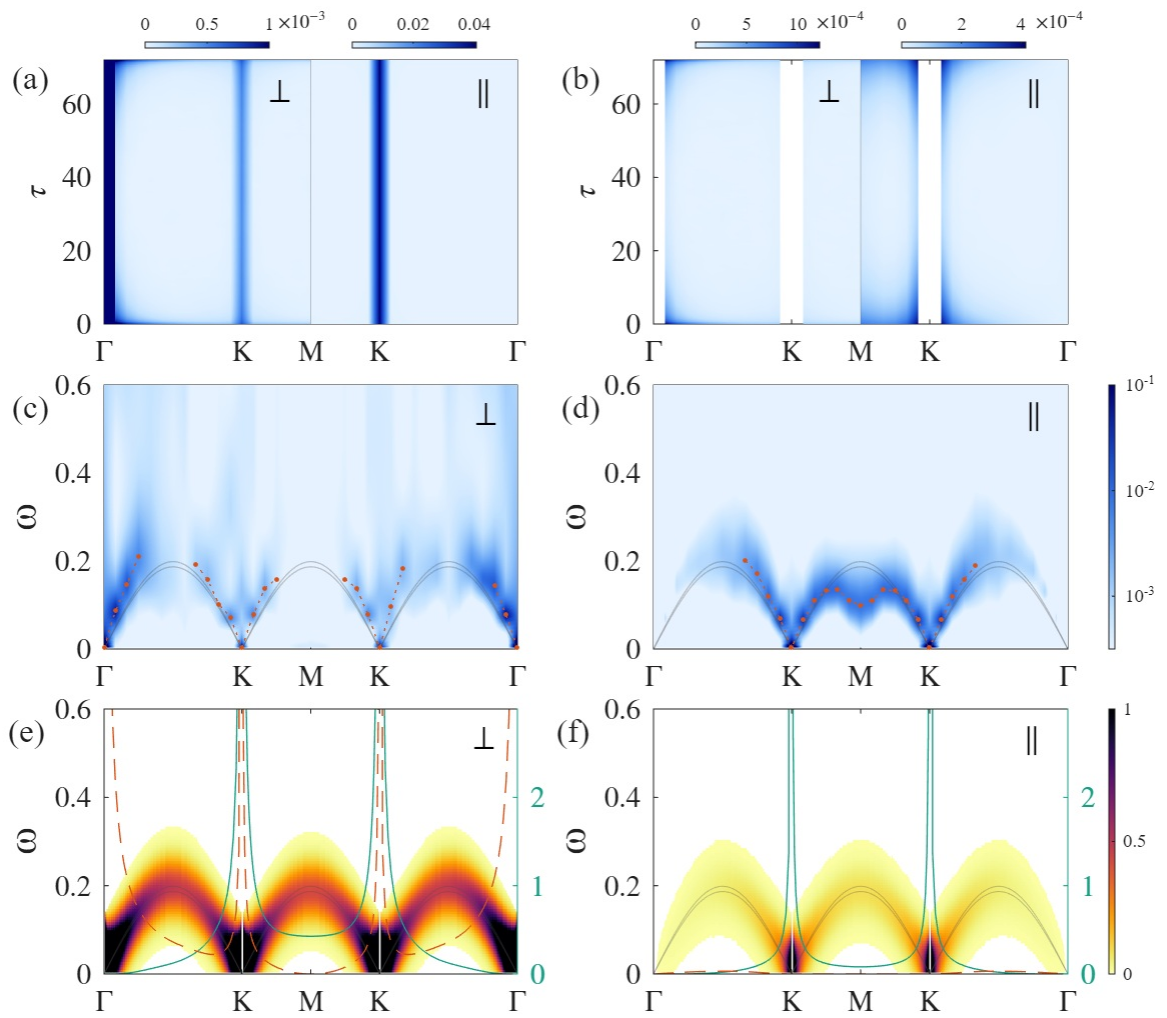} 
    \caption{(a) Imaginary-time DSF of the supersolid phase. (b) Replot of (a) highlighting the small values at momenta other than the $\Gamma$ and $K$ points. Real-frequency DSFs $S_{\parallel}(\mathbf{k},\omega)$ (c) and $S_{\perp}(\mathbf{k},\omega)$ (d) obtained via numerical analytic continuation. (e) and (f) are the corresponding spectra calculated using LSWT. The solid circles in (c) and (d) denote the peak positions of the QMC spectra, while the black solid lines represent the LSWT magnon dispersions. The green solid and red dashed lines in (e) and (f) indicate the spectral-weight prefactors in LSWT. The parameters are $t/V=0.06$ and $\mu/V=3$. }
    \label{fig3}
\end{figure}

\textit{\textcolor{blue}{Dynamical signatures of the supersolid phase.--}}
Figure~\ref{fig3} illustrates the longitudinal and transverse dynamical correlations of the supersolid phase at $t/V=0.06$ and $\mu/V=3$.
In both the imaginary-time and real-frequency domains, the two components display consistently distinct momentum distributions: $S_{\perp}$ is primarily distributed around the $\Gamma$ and $K$ points, whereas $S_{\parallel}$ dominates along the $K-M$ high-symmetry path.
This distinct momentum-space separation highlights the anisotropic nature of the magnetic excitations, which is also qualitatively captured by LSWT through the prefactors of the $\delta$ function (see SM).


The contour of $S_{\parallel}(\boldsymbol{k},\tau)$ features a persistent spectral line at the $K$ point, which corresponds to a dominant spectral weight at $\omega=0$, marking the static diagonal order in the $z$ direction. Accompanying this is a rapidly decaying diffuse region-two orders of magnitude weaker in intensity-whose profile peaks at the $K$ and $M$ points and corresponds to the finite-energy dispersion relation. Consequently, in the frequency domain, $S_{\parallel}(\boldsymbol{k},\omega)$ reveals a Goldstone mode near the $K$ point, alongside a roton-like minimum at the $M$ point driven by a strong downward energy renormalization.

By contrast, $S_{\perp}(\mathbf{k},\tau)$ exhibits persistent spectral lines at both the $K$ and $\Gamma$ points, though the intensity at $K$ is significantly weaker. While the $\Gamma$-point line signifies the uniform superfluid order in the $xy$ plane, the $K$-point line highlights the sublattice-dependent nature of the superfluid component, associated with the underlying $z$-directional solid order. By directly reflecting the off-diagonal order and indirectly capturing the diagonal order, this unique dual feature in $S_{\perp}(\mathbf{k},\tau)$ unambiguously distinguishes the supersolid from the pure superfluid phase (which features only a $\Gamma$-point line) and the pure solid phase (which exhibits no spectral line).
Similarly, the surrounding diffuse contour near $\tau=0$ yields linear gapless Goldstone modes in $S_{\perp}(\mathbf{k},\omega)$ around the $\Gamma$ and $K$ points. 
Moving away from these high-symmetry points, the spectral peaks systematically broaden into a diffuse distribution, which corresponds to a faster decay in imaginary time $\tau$. Notably, the spectrum also hosts a high-energy continuum extending above the primary Goldstone branches.
Furthermore, $S_{\perp}(\mathbf{k},\omega)$ features a secondary peak at the $K$ point around $\omega=0.117 (\approx  J_x)$, potentially indicating another pseudo-Goldstone mode~\cite{Gao2024Double}. However, its precise resolution is hindered by a limited spectral weight and the overlap with the high-energy continuum.


While two low-energy branches exist based on LSWT, their manifestations in $S_{\parallel}$ and $S_{\perp}$ are strongly momentum-selective-a feature qualitatively captured by the LSWT prefactors. The lowest-energy branch concentrates at the $K$ points for both components [green lines, Figs.~\ref{fig3}~(e) and (f)], whereas the second branch is exclusively transverse, peaking at both $\Gamma$ and $K$ [red dashed lines]. This mode distribution naturally dictates that $S_{\parallel}(\mathbf{k},\omega)$ exhibits solely a Goldstone mode at $K$, while $S_{\perp}(\mathbf{k},\omega)$ hosts a single Goldstone mode at $\Gamma$ and a coexistence of both modes at the $K$ point.

Although LSWT captures the dynamical signatures qualitatively, the true nature of the excitations goes beyond this semi-classical picture. Excitation lineshapes deviate profoundly from the sharp $\delta$-function quasiparticle peaks predicted by LSWT; instead, QMC reveals broad, non-Lorentzian profiles with significant asymmetry and continuum contributions. In the spectral maps, this quasiparticle breakdown appears as diffuse intensity rather than coherent branches. Moreover, this broadening is distinctly momentum-dependent: the spectrum at the $M$ point is more diffuse than at the $K$ point, exhibiting a larger linewidth and a pronounced high-energy tail. Crucially, LSWT fails entirely to capture the roton minimum, erroneously predicting a saddle point at $M$.

\textit{\textcolor{blue}{Dynamical evolution across the supersolid phase transitions.--}}
We next investigate the evolutions of dynamical correlations across the supersolid-superfluid and supersolid-solid phase transitions, thereby establishing robust signatures that uniquely characterize the supersolid phase. At fixed $\mu/V=3$, increasing $t/V$ drives the system from the supersolid to the superfluid phase. In the supersolid regime, $S_{\parallel}(\mathbf{k},\omega)$ exhibits a gapless Goldstone mode at the $K$ point, whose velocity increases monotonically with $t/V$, reflecting enhanced phase stiffness. Meanwhile, the roton feature at the $M$ point is progressively suppressed. Upon entering the superfluid phase, the roton mode vanishes; the dispersion near $M$ evolves into a local maximum consistent with LSWT predictions, while a finite gap opens at the $K$ point, signaling the breakdown of crystalline order. In contrast, for the transverse component $S_{\perp}(\mathbf{k},\omega)$, the spectral weight at finite momenta is continuously suppressed and vanishes completely upon entering the superfluid phase. In the imaginary-time domain, the supersolid--superfluid transition is characterized by the disappearance of the spectral line at the $K$ point in both components.


Along the cut toward the $\rho=2/3$ solid at fixed $t/V=0.06$, the spectral weight of $S_{\parallel}(\mathbf{k},\omega)$ progressively concentrates at the $K$ and $\Gamma$ points while being suppressed elsewhere, consistent with the development of crystalline order. Correspondingly, the spectral lines of $S_{\parallel}(\mathbf{k},\tau)$ are progressively enhanced at the $K$ and $\Gamma$ points. Deep in the solid phase, the spectral weight is almost entirely concentrated at these two momenta, with the intensity at the $K$ point being significantly stronger than that at the $\Gamma$ point.



Concurrently, the $S_{\perp}(\mathbf{k},\tau)$ spectral lines at $\Gamma$ and $K$ are continuously suppressed, transferring their intensity to other momenta. This redistributes the low-energy spectral weight in $S_{\perp}(\mathbf{k},\omega)$ to previously unoccupied wave vectors. Upon entering the solid phase, the imaginary-time spectral lines vanish as the weight spreads across the Brillouin zone, yielding a fully gapped dynamical spectrum. Although LSWT qualitatively captures the dispersions, the higher-energy branch centered near $K$ prominently displays continuum features [see Fig.~\ref{fig2}(d)].

\begin{figure}[htp]
    \centering
    \includegraphics[width=0.5\textwidth]{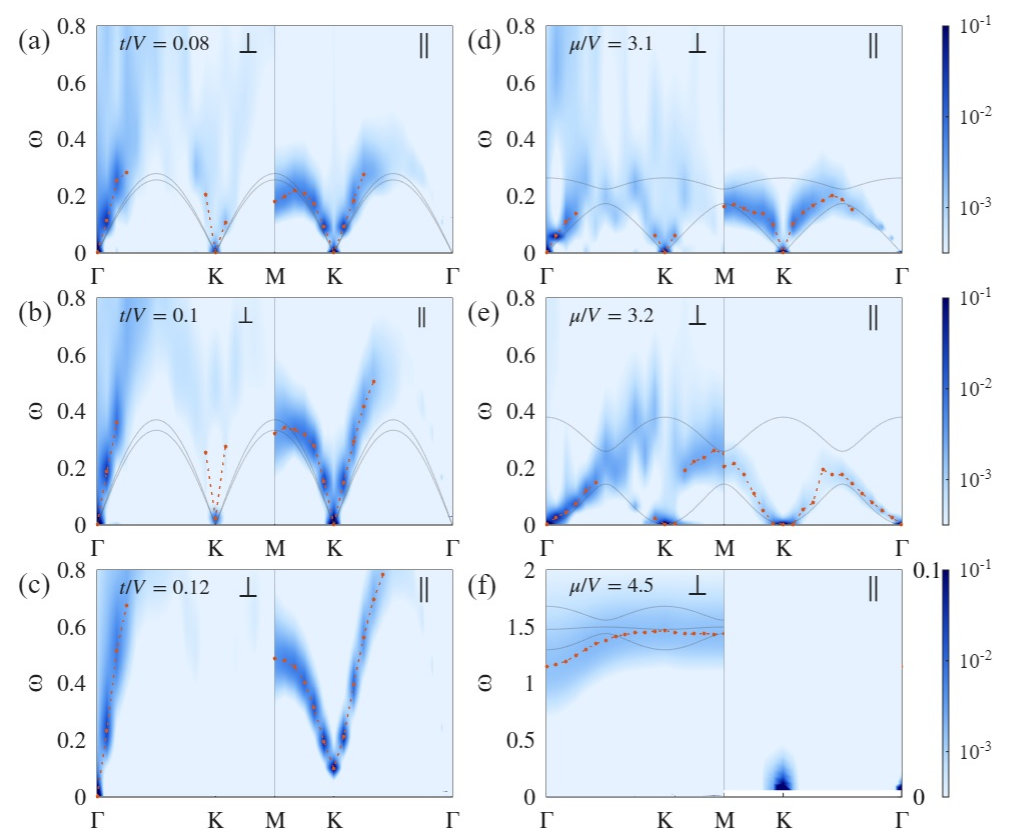} 
    \caption{Evolution of the dynamical spectra across the phase transitions. Evolution from supersolid to superfluid at fixed $\mu/V=3$ with the critical point $t/V=0.115$: (a) $t/V=0.08$, (b) $t/V=0.1$, and (c) $t/V=0.12$. Evolution from supersolid to solid at fixed $t/V=0.06$ with the critical point $\mu/V=3.26$, with (d) $\mu/V=3.1$, (e) $\mu/V=3.2$, and (f) $\mu/V=4.5$.}
    \label{fig4b}
\end{figure}

\textit{\textcolor{blue}{Conclusions.--}}
Combining LSWT, QMC imaginary-time correlations $S(\mathbf{k},\tau)$, and real-frequency spectra $S(\mathbf{k},\omega)$ establishes the defining dynamical signatures of the hardcore-boson supersolid. The longitudinal spectrum $S_{\parallel}(\mathbf{k},\omega)$ reveals a gapless Goldstone mode at $K$ and a roton minimum at $M$. Meanwhile, the transverse spectrum $S_{\perp}(\mathbf{k},\omega)$ hosts a primary gapless Goldstone mode at $\Gamma$, alongside a minor Goldstone component and a probable pseudo-Goldstone mode at $K$. Some of the dynamical features are directly mirrored in the imaginary-time domain, where the supersolid is uniquely characterized by the coexistence of a strong $\Gamma$-point and a weaker $K$-point spectral line in $S_{\perp}(\mathbf{k},\tau)$. 

The determined dynamical signatures of bosonic supersolids can provide direct insights into spin supersolids, for which unbiased large-scale simulations are lacking. Specifically, the superfluid-related Goldstone mode in spin supersolids is expected to shift to the K point. Moreover, the robust emergence of an M-point roton minimum in both supersolids, despite differing underlying configurations, may constitute a universal signature of supersolidity. Collectively, these predictions may guide the interpretation of recent inelastic neutron-scattering experiments on spin supersolid in triangular quantum magnets.



\textit{\textcolor{blue}{Acknowledgments.--}}The authors thank X.-C. Wang, Y. Gao, W. Li, H.-J. Liao, H. Shao, W.-A. Guo for helpful discussions. H.G. acknowledges support from the NSFC grant No.~ 12574249 and the BNLCMP open research fund under Grant No. 2024BNLCMPKF023. X.Z. acknowledges support from the NSFC grant No. 12304177 and the Natural Science Foundation of Jiangsu Province under Grant BK20230907. RTS is supported by the grant DOE DE-SC0014671 funded by
the U.S. Department of Energy, Office of Science. S.F. is supported by the National Key Research and Development Program of China under Grant Nos. 2023YFA1406500 and 2021YFA1401803, and NSFC under Grant No. 12274036. 

\bibliographystyle{apsrev4-1}
\bibliography{ref}

\end{document}